\documentclass[aps,pre,twocolumn,preprintnumbers,amsmath,amssymb]{revtex4}
\usepackage{graphicx,epsfig}
\usepackage{dcolumn}
\usepackage{wrapfig}
\usepackage{bm}
\usepackage{color}

\begin{document}

\title{Towards active microfluidics: Interface turbulence in thin liquid films with floating molecular machines }
\date{\today}

\author{Sergio Alonso$^{1,2}$ and Alexander S. Mikhailov$^{1}$}

\affiliation{
$^{1}$ Abteilung Physikalische Chemie, Fritz-Haber-Institut der
Max-Planck-Gesellschaft, Faradayweg 4-6, 14195 Berlin, Germany.\\
$^{2}$ Physikalisch-Technische Bundesanstalt, Abbestrasse 2-12, 10587 Berlin,
Germany. }

\begin{abstract}

Thin liquid films with floating active protein machines are considered. Cyclic
mechanical motions within the machines, representing microscopic swimmers,
lead to molecular propulsion forces applied to the air-liquid interface. We
show that, when the rate of energy supply to the machines exceeds a threshold,
the flat interface becomes linearly unstable. As the result of this
instability, the regime of interface turbulence, characterized by irregular
traveling waves and propagating machine clusters, is established. Numerical
investigations of this nonlinear regime are performed. Conditions for the
experimental observation of the instability are discussed.

\end{abstract}

\maketitle

\section{Introduction}

Molecular machines are protein molecules which can transform chemical energy
into ordered internal mechanical motions. The classical examples of such
machines are molecular motors kinesin and myosin, where internal mechanical
motions are used to transport cargo along microtubules and filaments. Many
enzymes operate as machines, using internal conformational motions to
facilitate chemical reactions. Other kinds of machines, operating as ion pumps
or involved in genetic processes, are also known. Moreover, artificial
nonequilibrium nanodevices, similar to protein machines, are being developed \cite{kay2007}.

The cycle of a protein machine is typically powered by chemical energy
supplied with an ATP molecule. When an ATP molecule binds to a protein
machine, the initial protein-ATP complex is out of equilibrium and the process
of ordered conformational relaxation of this complex towards its equilibrium
state starts. When this state is reached, ATP is converted into ADP and this
product molecule leaves the complex. The emerging free protein molecule is out
of equilibrium and another conformational relaxation process, returning the
free protein to its original equilibrium state begins. Thus, the cycle
consists of two conformational motions. It is important to note that the
forward and back conformational motions inside each cycle do not coincide -
they correspond to relaxation processes in two different physical objects,
i.e. the free protein and the protein-ATP complex  \cite{togashi2007}.

Recently, much attention has been attracted to microscale swimmers operating
at low Reynolds numbers. Generally, it can be shown that any physical object,
cyclically changing its shape in such a way that the internal forward and back
motions are different, propels itself through the liquid  \cite{purcell1976,shapere1987}. Elementary models of such
swimmers, constructed by joining together mobile links \cite{purcell1976,becker2003} or by connecting a few spheres by several 
actively
deformable links \cite{najafi2004,earl2007,golestanian2008}, have been considered. Typically,
the concept of molecular swimmers is applied to explain active motion of
bacteria and other microorganisms. However, individual macromolecules that
operate actively as machines are also capable of self-motion \cite{kay2007,golestanian2008prl}

If a swimmer is attached to some support, preventing its translational motion,
it exhibits force acting on the mechanical support. The generated force
is equal to the stall force which needs to be applied to a swimmer to prevent
its translational motion. When many swimmers are attached to a distributed
support, their collective operation produces mechanical pressure acting on the
support. This situation is, for example, encountered for molecular machines
representing active protein inclusions (such as ion pumps) in biological
membranes. It has been shown that the combined effects of the mechanical
forces generated by the machines and their lateral motions inside the membrane
can lead to membrane instabilities \cite{Ramaswamy2000}.

\begin{figure}[b]
\begin{center}
\epsfig{file=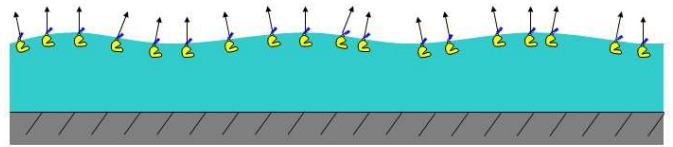, width=3.5in}
\end{center}
\caption{(Color online) Sketch of a thin film with floating molecular machines generating propulsion forces.}
\label{fig1}
\end{figure}

The aim of this paper is to investigate a different situation involving
molecular machine populations. We consider hypothetical machines which are
floating on top of a thin liquid layer and thus represent \textit{active
surfactants} (Fig.\ref{fig1}). Because the considered surfactant molecules are
microswimmers, they generate local pressure applied to the liquid-air
interface and proportional to their local concentration. Additionally, they
are subject to surface diffusion and, as surfactants, also modify local
surface tension. Our main analytical result, supported by numerical
simulations, is that, if the surface density of machines is sufficiently high
and if the rate of supply of chemical energy to the machine population exceeds
a threshold, the equilibrium flat interface becomes unstable and
hydrodynamical flows inside the liquid layer spontaneously develop. This
instability is accompanied by spatial redistribution of floating machines and
their collective motions over the surface. It leads to the emergence of a
special kind of surface turbulence. Such phenomena can provide the basis for
active microfluidics, where hydrodynamical motions in liquid layers are
induced and controlled by molecular machines located at the surface.

The next two sections are devoted to the derivation and the stability analysis
of a model of such molecular swimmers attached to the surface of a thin liquid
film. A collection of numerical results is presented in section IV. Finally,
a model and numerical results are discussed, and some estimates of the forces of
such molecules are given. The possible experimental
realization is also discussed.

\section{Formulation of the Model}

Population of identical molecular machines floating on the surface of a thin
liquid film will be considered. These machines perform cycles of
conformational changes (with a characteristic time $t_{c}$) enabled by the
supply of energy through ATP molecules present in the liquid. We are
interested in macroscopic collective effects and do not specify the details of
machine operation. It will be assumed that, on the average, each machine
generates the force which is applied to the air-liquid interface. This average
force acts along the normal direction (the forces in the lateral direction
vanish after averaging because of the rotational diffusion). Each machine
cycle is initiated by binding of an ATP molecule and the average force is
proportional to the cycle frequency. Assuming that binding of ATP molecules
follows the Michaelis-Menten law, the average force per machine molecule is

\begin{equation}
f=f_{0}\frac{[ATP]}{K_{ATP}+[ATP]}. \label{ATP}%
\end{equation}
Here, $f_{0}$ is the maximal force under the ATP saturation conditions,
$\left[  ATP\right]  $ denotes the ATP concentration in the liquid, and
$K_{ATP}$ is the characteristic concentration at which saturation begins.

Because of the cycles, the floating machines produce additional pressure
acting on the air-liquid interface. This pressure $p_{m}$ is proportional to
the local surface concentration of the machines and the average force
generated by an individual machine, i.e. $p_{m}=f_{m}c$, where $f_{m}$ is the
force per Mol of molecules ($f_{m}=f N_{A}$ where $N_{A}$ is the Avogadro
number). Note that the machines are still sufficiently well separated one from
another and possible synchronization effects of their cycles (cf. \cite{casagrande2007}) are therefore neglected.

Before proceeding to the detailed formulation of the model, we want to outline
the origin of the expected surface instability. Suppose that the average
force, generated by machines is directed upwards and therefore the machines
are pulling the liquid up in the vertical direction (Fig. 2A). If machine
concentration is increased in some region, the pulling pressure is higher in
this region, inducing local rise in the liquid film thickness. This however
leads to lateral hydrodynamical flows which are directed inwards and bring
even more machines into the region. As a result, the positive feedback,
responsible for the instability, is established. Note that surface diffusion
of floating machines and capillary forces are acting in the opposite
direction, suppressing the instability. Thus, it is observed only if the
average force $f$ generated by a machine is large enough, setting a threshold
in terms of the energy supply rate (i.e., of the ATP concentration in the
solution). The instability is not possible when the force is
downwards directed, thus inducing local depressions of the liquid layer (Fig. 2B), Then, in contrast to the 
previous case, hydrodynamical flows remove machines
from the depression, restoring the equilibrium flat film.

\begin{figure}[t]
\begin{center}
\epsfig{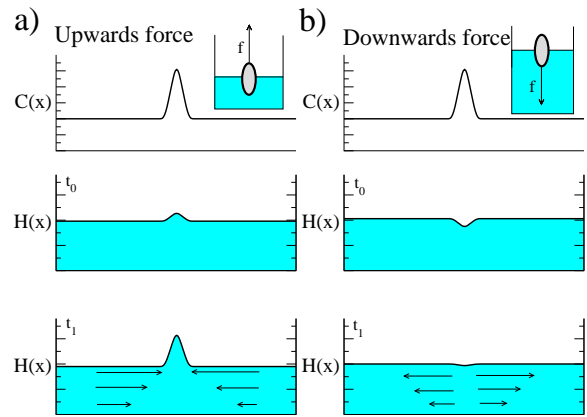}
\end{center}
\caption{(Color online) Mechanisms of (a) instability development for upwards directed propulsion forces, and of 
(b) instability damping for downwards directed propulsion forces.}
\label{figIO}%
\end{figure}

Temporal evolution of the surface concentration field $c$ is described by the
equation :

\begin{equation}
\partial_{t}c+\vec{\nabla}\left(  \vec{u}c\right)  =d\nabla^{2}c,
\end{equation}
where $\vec{u}$ is the lateral fluid velocity and $d$ is the surface diffusion.

Floating proteins represent surfactants and reduce the surface tension of the
interfaces (see \cite{beverung1999}). To take this effect into account, we
assume that the surface tension coefficient $\sigma$ decreases linearly with
the machine concentration,%

\begin{equation}
\sigma=\sigma_{0}-\sigma_{c}c\text{.}%
\end{equation}

Hydrodynamical flows, induced by the gradients in machine concentration,
should be further considered. We assume that the liquid layer is so thin that
the lubrication approximation, typically employed in microfluidics
\cite{oron1997}, is justified. As shown in Appendix A, the evolution equation
for the local hight $h$ of the interface has the form

\begin{equation}
\partial_{t}h=\frac{1}{\mu}\vec{\nabla}\left(  \frac{h^{3}}{3}\vec{\nabla
}p-\frac{h^{2}}{2}\vec{\nabla}\sigma\right)  ,
\end{equation}
where the local pressure is $p=-\sigma\nabla^{2}h-p_{m}$ and $\mu$ is the
viscosity of the fluid. Determining the lateral flow velocity at the
interface (see Appendix A) and substituting it into the evolution equation for the surface
concentration, a closed set of two partial differential equations is obtained.

Explicitly, the considered dynamics of thin films with active surfactants is
described by equations:

\begin{align}
\partial_{t}h  &  =-\frac{1}{\mu}\vec{\nabla}\left[  \frac{h^{3}}{3}\left(
\vec{\nabla}\left(  \sigma_{0}\nabla^{2}h-\sigma_{c}c\nabla^{2}h+f_{m}%
c\right)  \right)  \right] \nonumber\\
&  +\frac{1}{\mu}\sigma_{c}\vec{\nabla}\left(  \frac{h^{2}}{2}\vec{\nabla
}c\right)  , \label{dth}%
\end{align}

\begin{align}
\partial_{t}c  &  =-\frac{1}{\mu}\vec{\nabla}\left[  c\frac{h^{2}}{2}%
\vec{\nabla}\left(  \sigma_{0}\nabla^{2}h-\sigma_{c}c\nabla^{2}h+f_{m}%
c\right)  \right] \nonumber\\
&  +\frac{1}{\mu}\sigma_{c}\vec{\nabla}\left(  ch\vec{\nabla}c\right)
+d\nabla^{2}c. \label{dtc}%
\end{align}

Note that we have assumed that the active surfactant is insoluble \cite{oron1997,dewit1994}.

For subsequent analysis, it is convenient to write these equations in the
dimensionless form. The (equilibrium) liquid layer thickness $h_{0}$ will be
used as the length unit, time will be measured in units of $\mu h_{0}
/\sigma_{0}$, and local concentration $c$ in units of the equilibrium machine
concentration $c_{0}$. Changing the variables as $T=(\sigma_{0}/\mu h_{0})t$,
$H=h/h_{0}$, $X=x/h_{0}$ and $C=c/c_{0}$, we obtain
\begin{align}
\partial_{T}H  &  =-\frac{1}{3}\vec{\nabla}\left(  H^{3}\vec{\nabla}\left[
\left(  1-AC\right)  (\nabla^{2}H)+BC\right]  \right) \nonumber\\
&  +\frac{A}{2}\vec{\nabla}\left(  H^{2}(\vec{\nabla}C)\right)  , \label{pth2}%
\end{align}

\begin{align}
\partial_{T}C  &  =-\frac{1}{2}\vec{\nabla}\left[  CH^{2}\vec{\nabla}\left(
\left(  1-AC\right)  (\nabla^{2}H)+BC\right)  \right] \nonumber\\
&  +A\vec{\nabla}\left[  CH\vec{\nabla}C\right]  +D\nabla^{2}C. \label{ptc2}%
\end{align}

The new model equations include only three dimensionless parameters

\[
A=\frac{\sigma_{c}c_{0}}{\sigma_{0}},\qquad B=\frac{f_{m}c_{0}h_{0}}%
{\sigma_{0}},\qquad D=\frac{d\mu}{\sigma_{0}h_{0}}.
\]
The parameter $A$ specifies the characteristic strength of floating machines
as the surfactant species (decreasing the local surface tension of the
interface). The parameter $B$ specifies the magnitude of the pressure
generated by the cycling molecular machines; it is controlled by the rate of
energy supply to the system. Finally, $D$ is the dimensionless diffusion
coefficient of floating machines.

Comparing the last terms in equation (\ref{ptc2}), we notice that
spreading of floating machines, induced by changes in the surface tension, has
the same functional form as surface diffusion. If relative variations of the
film thickness and machine concentration are small ($H\sim C\sim1$), the
effective diffusion coefficient of this process is $A$. Our estimates below in
Sect. V indicate that the genuine diffusion constant $D$ of floating machines
is typically much smaller than the effective diffusion constant $A$. Having
this in mind, we retain the terms including the coefficient $D$ in our
analytical investigations, but put $D=0$ when numerical simulations of the
model are performed. Results with non-zero diffusion coefficient have been 
also obtained but they are very close to the results with $D=0$.

As shown in Appendix A, the dimensionless horizontal ($\vec{U}$) and vertical
($W$) velocities of the liquid film at hight $Z=z/h_{0}$ and horizontal
spatial location $X$ can be found as%

\begin{align}
&  \vec{U}(X,Z)=-A\vec{\nabla}CZ\\
&  +\frac{1}{2}\left(  B\vec{\nabla}C+\vec{\nabla}\left(  \left(  1-AC\right)
\nabla^{2}H\right)  \right)  (2HZ-Z^{2}),\nonumber
\end{align}

\begin{align}
&  W(X,Z)=-\frac{1}{2}A\nabla^{2}CZ^{2}\\
&  +\frac{1}{2}\left(  HZ^{2}-\frac{Z^{3}}{3}\right)  \nabla^{2}\left(
BC+\left(  1-AC\right)  \nabla^{2}H\right) \nonumber\\
&  -Z^{2}\vec{\nabla}H\left(  (B-A\nabla^{2}H)\vec{\nabla}C+(1-AC)\vec{\nabla
}\nabla^{2}H\right) \nonumber
\end{align}
when the fields $C$ and $H$ are known. Both velocities vanish at the bottom of
the liquid film, at $Z=0$, because of the no-penetration and no-slip boundary
conditions imposed there.

\section{Linear stability analysis}

The model always has the stationary uniform state $C=H=1$. To perform the
linear stability analysis of this state, small perturbations $H=1+\Delta H$
$\exp(iKX+ST)$ and $C=1+\Delta C$ $\exp(iKX+ST)$ in the form of plane waves
with the wavenumber $K$ are introduced. Linearizing evolution equations with
respect to small perturbations $\Delta H$ and $\Delta C$ and solving the
linearized equations, growth rates $S=S(K)$ of such perturbations are obtained,

\begin{align}
S(K) &  =-\frac{1}{2}\left(  A-\frac{B}{2}+D\right)  K^{2}-\frac{1}
{6}(1-A)K^{4}\\
& \pm\frac{K^{2}}{2} ( \left[  \frac{1}{3}(1-A)K^{2}-A+\frac{B}
{2}-D\right]  ^{2}    \nonumber\\ 
&  + 2 \left(  1-A\right)  \left(  \frac{A}{2}-\frac{B}
{3}\right)  K^{2} )^{1/2}  .\nonumber
\end{align}

It can be easily checked that, in absence of the energy supply ($B=0$), both
rates are real and negative, so that the equilibrium flat film is stable as
should be expected. When the parameter $B$ is increased, the instability
develops at $B=B_{c}$ where
\begin{equation}
B_{c}=2(A+D). \label{boundary}%
\end{equation}
Above the instability threshold, traveling plane waves with the wavenumbers
$K$ near $K=K_{c}(B)$,
\begin{equation}
K_{c}(B)=\frac{\sqrt{3}}{2} \frac{\sqrt{B-B_{c}}}{1-A}, \label{wavenumber}%
\end{equation}
and the frequencies near $\Omega=\Omega_{c}(B),$
\begin{equation}
\Omega_{c}(B)\approx \frac{3}{16}\sqrt{2B-3A} \frac{ ~\left(B-B_{c}\right)^{3/2}}{1-A},
\label{frequency}%
\end{equation}
are growing.The fastest growing mode with $K=K_{c}$ and $\Omega=\Omega_{c}$ is
characterized by the growth rate%
\begin{equation}
\operatorname{Re}\left[  S_{c}(B)\right]  = \frac{3}{32}  \frac{~\left(  B-B_{c}\right)^{2}}{1-A}. \label{rate}%
\end{equation}

\begin{figure}[t]
\begin{center}
\epsfig{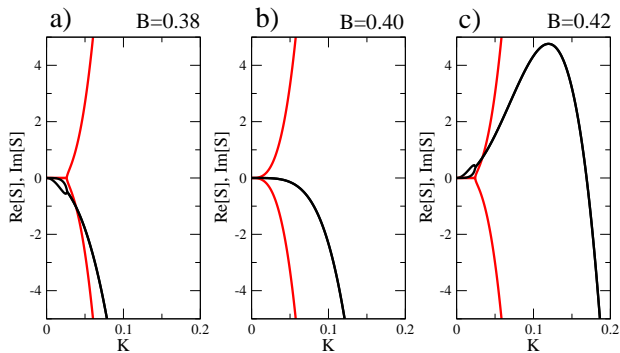}
\end{center}
\caption{(Color online) Real (solid line) and imaginary (light line) parts of the growth rate $S$ as 
functions of the
wavenumber $K$ of the perturbation for (a) $B=0.38,$ (b) $B=0.4$ and (c)
$B=0.42$. Other parameters are $A=0.2$ and $D=0$.}
\label{fig3}
\end{figure}

Figure 3 shows dependences $\operatorname{Re}\left[  S(K)\right]  $ and
$\operatorname{Im}\left[  S(K)\right]  $ at three values of the parameter $B$
below and above the instability boundary. While the fastest growing mode at
$B>B_{c\text{ }}$is always oscillatory, with $\operatorname{Im}\left[
S(K)\right]  \neq0$, above the instability boundary the system also always has
some standing growing modes with the wavenumbers $K$ smaller than $K_{c}$. As
the boundary is approached from above, both the wavenumber $K_{c}$ and the
frequency $\Omega_{c}$ decrease and vanish at $B=B_{c}$. 
The region with two real modes $S_1(K)$ and $S_2(K)$ always lies below $K_c$; it shrinks and vanishes at $B = B_c$. Similar
long-wavelength instabilities have previously been discussed for other
conservative systems (see \cite{cross1993}).

\section{Numerical Investigations of the Nonlinear Regime}

To investigate the behavior of the system in the nonlinear regime above the
instability onset, numerical simulations have been performed.
Equations(\ref{pth2},\ref{ptc2}) were integrated using the semi-implicit
method (see appendix B) for a one-dimensional system using periodic boundary
conditions. As the initial condition, the flat interface with a uniform
machine distribution was chosen and small random initial perturbations were applied.

\begin{figure*}[]
\begin{center}
\epsfig{file=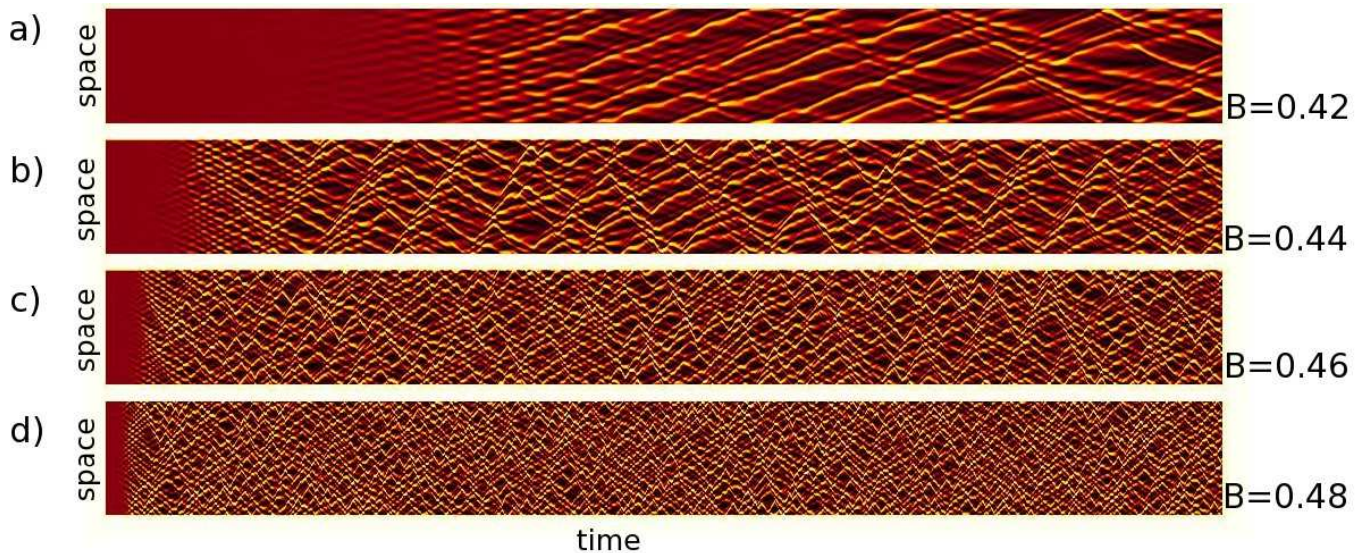, width=7.4in}
\end{center}
\caption{(Color online) Interface turbulence. Space-time diagrams display the
local film thickness $H$ (thicker regions correspond to bright and thinner regions to dark colors) depending on the spatial 
coordinate 
and time for : (a) $B=0.42$, (b) $B=0.44$, (c) $B=0.46$ and (d) $B=0.48$.
Other system parameters are $A=0.2$ and $D=0$. Numerical integrations for a
one-dimensional system of length $L_{0}=512$ (with $\Delta x=1$) and the total time interval of
$T_{0}=5 \cdot 10^5$ (with $\Delta t=0.01$).}
\label{fig4}
\end{figure*}

Our main observation is that the instability development results in the
emergence of a complex spatiotemporal regime which can be described as
\textit{interface turbulence}. This regime is characterized by spontaneous
appearance of \textit{traveling machine clusters} (i.e., of spatial regions
where the local machine concentration is increased) and of the accompanying
local interface modulations.

Figure(\ref{fig4}) gives an illustration of the turbulent regimes observed at
different different distances from the instability threshold. The local film
thickness $H$ is displayed here in gray scale depending on the spatial
coordinate (the vertical axis) and time. In the slightly supercritical regime
at $B=0.42$, the transient development of standing waves is first observed,
which is then followed by the emergence of an irregular pattern of traveling
and colliding waves. At larger deviations from the critical point, the
transients are faster and the irregular wave dynamics appears soon after the
instability onset. The characteristic spatial scale of the turbulence
decreases with the control parameter $B$, consistent with the predictions of
the linear stability analysis. The characteristic velocity of traveling waves
is also growing with $B$.

\begin{figure}[b]
\begin{center}
\epsfig{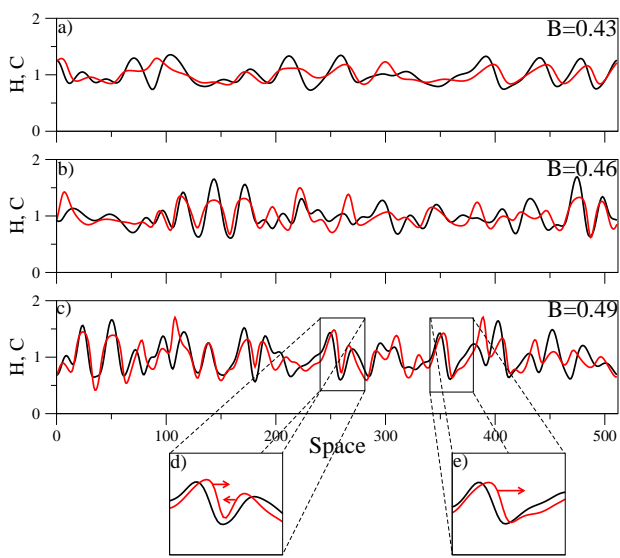}
\end{center}
\caption{(Color online) Snapshots of the concentration distributions (light
lines) and of the respective interface profiles (black lines) for (a)
$B=0.43$, (b) $B=0.46$ and (c) $B=0.49.$ Other parameters are $A=0.2$ and
$D=0$. Insets on the bottom show the collision of two waves (d) and the motion of a single wave (e). 
Arrows in the insets indicate the directions of motion of the distribution maxima.}
\label{fig5}
\end{figure}

Figure \ref{fig5} shows snapshots of computed turbulent patterns at different
deviations from the instability threshold.
Both the interface profiles and the corresponding concentration distributions
are presented here. Again, a decrease of the characteristic wavelength of the
irregular spatiotemporal patterns under an increase of the control parameter
$B$ can be noticed. Moreover, we see that the characteristic amplitude of the
waves grows with $B$. To illustrate the directions of wave propagation, arrows
are placed in the insets of this figure, where examples of colliding and traveling waves are
shown. These arrows indicate the directions in which the
respective concentration and interface profile maxima are shifting at the next
time moment. Note that the motions of machine clusters (i.e, of the local
concentration maxima) are typically guiding the motions of surface bumps.

\begin{figure}[t]
\begin{center}
\epsfig{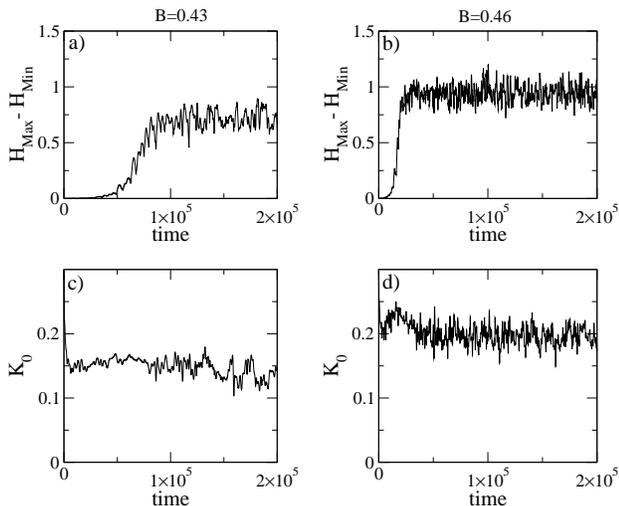}
\end{center}
\caption{Initial evolution of the pattern amplitudes $\Delta H(T)$ and of the
characteristic wavenumbers $K_{0}$ of the developing patterns for two
different values $B=0.43$ and $B=0.46$ of the control parameter. Other
parameters are $A=0.2$ and $D=0$. The system size is $L_{0}=512.$}
\label{fig6}
\end{figure}

Temporal transients leading to turbulent patterns are characterized in Fig. \ref{fig6}.
To construct it, we have determined the maximum and the minimum values
$H_{\max}$ and $H_{\min\text{ }}$of the film thickness as function of time,
starting from the initial moment. Their difference $\Delta H(T)=H_{\max
}(T)-H_{\min}(T)$ (where $T$ is the dimensionless time) can be chosen to describe 
the amplitude of the developing
patterns. As seen in the two top panels in Fig. \ref{fig6}, this amplitude first grows
(exponentially) and the undergoes saturation. The transient is shorter for the
larger deviation from the threshold ($B=0.46$) and the final mean amplitude of
the turbulent pattern is also then larger.

To estimate the characteristic wavenumber $K_{0}(T)$ of the developing spatial
patterns, their Fourier transforms were computed and the positions of dominant
maxima in the spatial power spectra were determined at different time moments.
As seen in the two bottom panels in Fig. \ref{fig6}, these wavenumbers do not
significantly change during the transients. The characteristic wavelengths of
the developed turbulent patterns are therefore not much different from those
of the critical modes.

\begin{figure}[t]
\begin{center}
\epsfig{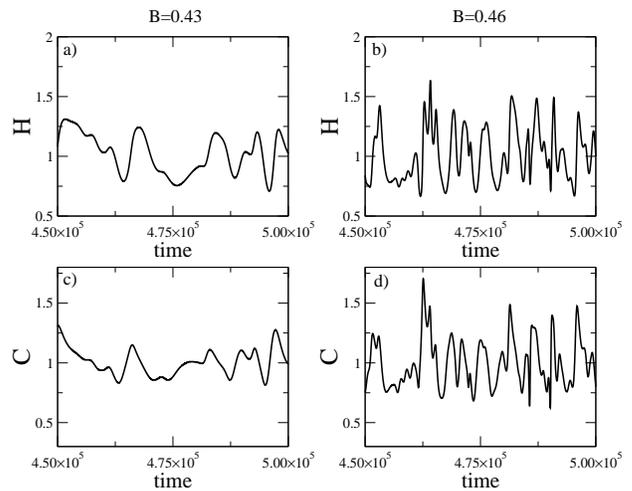}
\end{center}
\caption{Time dependences of the film hight $H(T)$ and concentration $C(T)$ at
a fixed spatial point for two different values of the control parameter $B$.
Other parameters and the system length are the same as in Fig.\ref{fig4}.}
\label{fig7}
\end{figure}

Figure \ref{fig7} shows temporal dependences of the hight $H(T)$ and the concentration
$C(T)$ at a fixed point of the system in the final turbulent state. The
characteristic time of the oscillations is clearly different for the two
values of the control parameter. Both properties fluctuate around some mean
values. The fluctuations are more rapid farther away from the instability threshold.

\begin{figure}[t]
\begin{center}
\epsfig{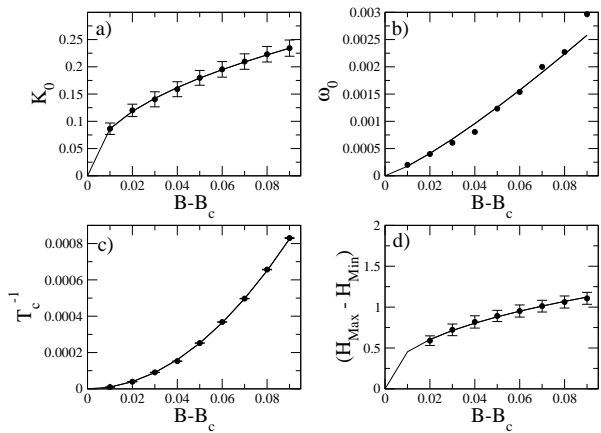}
\end{center}
\caption{Statistical properties of the interface turbulence at different
deviations from the instability threshold: (a) characteristic wavenumber,
(b) characteristic frequency, (c) characteristic inverse transient time and
(d) mean amplitude of the patterns. Dots show mean values, bars indicate
statistical dispersion of numerical data. Solid curves show power laws
corresponding to the linear stability analysis predictions (see the text). The
same system and simulation parameters as in Fig. 6.}
\label{fig8}
\end{figure}

Finally, Fig. \ref{fig8} displays the dependence of several selected statistical
properties of the final turbulent state on the deviation $\Delta B=B-B_{c}$
from the critical point $B_{c}$. The characteristic wavenumbers $K_{0}$ and
amplitudes $\Delta H$ have been computed as described above. Time averages of
these properties and statistical dispersion of the data are shown in Figs.
\ref{fig8}. The characteristic transient times $T_{c}$ (Fig. \ref{fig8}c) have been
estimated by fitting the initial computed time dependence for $\Delta H(T)$ to
the exponential law, $\Delta H(T)\sim\exp(T/T_{c})$.  
The characteristic frequency $\Omega_{0}$ of the patterns (Fig. \ref{fig8}d) is
estimated by computing their temporal power spectra and determining the
positions of the dominant maxima.

Solid curves in Fig. \ref{fig8} give fits of the simulation data to the power laws
predicted by the linear stability analysis. According to equations
(\ref{wavenumber}) and (\ref{frequency}), $K_{c}\sim$ $\Delta B^{1/2}$ and
$\Omega_{c}\sim\Delta B^{3/2}$. The transient time is determined by the rate
of growth as $T_{c}=1/\operatorname{Re}\left[  S_{c}(B)\right]  $ and,
according to equation (\ref{rate}), we expect that $T_{c}^{-1}\sim\Delta
B^{2}$. For the mean amplitude $\Delta H$ of the patterns, the quadratic fit
$\Delta H\sim\Delta B^{1/2}$ has been applied. We see that the statistical
properties of nonlinear patterns in the developed turbulence regime are still
in good agreement with the respective predictions based on the linear
stability analysis.

\section{Discussion and Conclusions}

Can predicted interface instabilities be experimentally observed? This
question has both chemical and physical aspects. On the chemical side, it is
known that proteins may indeed represent surfactants and thus float at the
air-water interface (see. e.g., \cite{beverung1999}). Although we cannot give
here a specific example, it seems plausible that some protein machines also
belong to this class. Moreover, other protein machines, including molecular
motors, can probably be made floating by chemical modification, i.e. by
attaching to them a hydrophobic group.

The physical question is whether the propulsion forces generated by individual
protein machines would be sufficient to induce the considered interface
instability. According to equation (\ref{boundary}), the instability is
reached when $B>B_{c}$ with $B_{c}=2(A+D)$. Taking into account the
definitions of dimensionless properties $A$, $B$ and $D$, the instability
condition implies that the propulsion force $f,$ generated by a single
machine, must exceed the threshold

\begin{equation}
f_{c}=\frac{2}{N_{A}}\left(  \frac{\sigma_{c}}{h_{0}}+\frac{d\mu}{c_{0}%
h_{0}^{2}}\right)  \label{force}%
\end{equation}
where $h_{0}$ is the film thickness, $c_{0}$ is the surface concentration of
floating machines, $d$ is their surface diffusion constant, $\mu$ is the
liquid viscosity, $N_{A}$ is the Avogadro number, and $\sigma_{c}$ specifies
the surfactant capacity of the considered biomolecules (the surface tension
coefficient $\sigma$ depends as $\sigma=\sigma_{0}-\sigma_{c}c$ on their
concentration $c$).

For numerical order-of-magnitude estimates, we choose $\mu\approx10^{-2}$ g
cm$^{-1}$s$^{-1}$ (the viscosity of water), $d\approx10^{-7}$ cm$^{2}$s$^{-1}$
(characteristic diffusion constant of large biomolecules in water solutions \cite{kim1992}).
Based on the experimental data given in Ref. \cite{beverung1999}, we take
furthermore $\sigma_{c}\approx10^7~g~cm^2~s^{-2}~mMol^{-1}$.

Both terms in equation (\ref{force}) are inversely proportional to the liquid
film thickness $h_{0}$ and the smaller critical forces are therefore expected
for the thicker layers. Note, however, that this result is based on the
lubrication approximation for thin liquid films, requiring that the film
thickness is much smaller than the characteristic wavelength of the flow
patterns. As we have seen, the characteristic wavelength of the interface
turbulence in the considered system depends on the distance from the critical
point (cf. equation (\ref{wavenumber})) and, in principle, can be arbitrarily
large sufficiently close to the instability threshold. Taking into account
experimental limitations making it very difficult to work too close to the
threshold, we choose however $h_{0}=1mm$ as the characteristic maximum film thickness.

Concentration $c_{0}$ of floating proteins enters only into the second term in
equation (\ref{force}). This term, depending on surface diffusion of floating
machines, can be neglected in comparison with the first term, if the condition
$c_{0}\gg c_{d}$ with
\begin{equation}
c_{d}=\frac{d\mu}{\sigma_{c}h_{0}} \label{diffusion}%
\end{equation}
holds. Substituting numerical values, we get $c_{d}=10^{-18}~Mol/cm^2$. Thus, already for
very low surface concentrations of proteins, effects of their diffusion can be
neglected in the considered problem.

Neglecting diffusion effects, the critical propulsion force of a single
machine is estimated as
\begin{equation}
f_{c}=\frac{2\sigma_{c}}{N_{A}h_{0}}. \label{force1}%
\end{equation}
Note that it does not depend on the protein concentration.

Substitution of the numerical values into equation (\ref{force1}) yields
$f_{c}=3\cdot10^{-5}~pN$. Can a single molecular machine generate the hydrodynamical
propulsion force of that magnitude?

Direct measurements of molecular propulsion forces are still not available. It
is known that molecular motors, such as myosin or kinesin, can generate
mechanical forces of about $1~pN$ \cite{Finer1994,fisher1999,fisher2001}, but
these data refers to the molecules moving along microtubules and filaments,
not to the swimmers. 

For a simple example, we consider the elementary three-body Purcell swimmer
\cite{purcell1976}. Its characteristic propulsion velocity is about $V=\Delta
L/\tau$, where $\Delta L$ is the displacement per single cycle and $\tau$ is
the characteristic time of the cycle (cf. \cite{becker2003,tam2007}). The
viscous friction force of an object of linear size $L$ that moves with
velocity $V$ through the liquid is by the order-of-magnitude $f_{0}=6\pi\mu
LV$. Thus,
propulsion at velocity $V=\Delta L/t_{c}$ would require the propulsion force
about $f_{0}=6\pi\mu L\Delta L/t_{c}$. Choosing $L\approx50$ nm, $\Delta
L=0.1L$ and $\tau=1$ ms and considering water as the liquid, we obtain for the
molecular propulsion force the rough estimate of $f_{0}=10^{-3}$pN. Similar
estimates are obtained for other known elementary swimmers. Note that the
actual average propulsion force $f$ of a machine additionally depends on the
frequency of machine cycles and the angle with respect the interface, therefore, according to equation (\ref{ATP}),
this gives the estimate of the maximum propulsion force under the
energy-supply saturation conditions and the most efficient orientation of the machine.

While molecular propulsion forces are quite small, according the above
estimates they would still be sufficient to induce the film instability and
lead to the interface turbulence. Therefore, we conclude that the experimental
observation of the predicted effects is \textit{principally possible}.

The experiments aimed at detecting instabilities of thin liquid layers induced
by floating actively operating machines would allow to directly estimate
actual propulsion forces generated by particular biomolecules. Investigations
of such nonequilibrium hydrodynamical systems would be very important from the
perspective of \textit{active microfluidics}, where active motions in thin
liquid layers are produced by floating biomolecular propellers.

Financial support from the EU Marie Curie RTN "Unifying principles in
nonequilibrium pattern formation" and from the German Science Foundation
Collaborative Research Center SFB 555 "Complex Nonlinear Processes" is
acknowledged. We are grateful to U. Steiner, D. Barbero and U. Thiele for
valuable discussions.

\appendix 

\section{Hydrodynamic equations}

We consider the situation when the film thickness $h_{0}$ is the smallest
characteristic length of the system. In this case, the lubrication
approximation can be used which corresponds to an expansion in the small
parameter $\epsilon=h_{0}/\lambda$, with $\lambda$ being the characteristic
wavelength of the patterns in the lateral direction. Hydrodynamical equations
in the lubrication approximation are simplified and take the form (see, e.g.,
\cite{oron1997})%

\begin{align}
&  \mu\partial_{zz}\vec{u}=\vec{\nabla}p,\nonumber\\
&  \partial_{z}p=0, \label{stokes}%
\end{align}
with the incompressibility condition
\begin{equation}
\vec{\nabla}\vec{u}+\partial_{z}w=0.
\end{equation}
Here, $w$ is the vertical component of the fluid velocity and $\vec{u}$ is its
horizontal component; $\vec{\nabla}$ is the differential operator acting only
on the horizontal coordinates. Note that, as follows from these equations,
pressure $p$ is constant along the vertical direction.

The boundary conditions at the bottom of the film, in contact with the solid
support, are
\[
w=\vec{u}=0\text{ at }z=0.
\]
At the free air-liquid interface $z=h$, the balance of horizontal and vertical
forces should separately hold, implying that%

\begin{align}
&  \mu\partial_{z}\vec{u}=\vec{\nabla}\sigma,\\
&  p=-\sigma\nabla^{2}h-p_{m}, \label{p}%
\end{align}
where $\sigma$ is the surface tension coefficient and $p_{m}$ is the
additional pressure produced by floating active molecules.

Conservation of the total film volume implies moreover that the equation.%

\begin{equation}
\partial_{t}h+\vec{\nabla}\left(  \int_{0}^{h}\vec{u}dz\right)  =0
\label{interface}%
\end{equation}
should hold.

Integrating equations.(\ref{stokes}) and taking into account the boundary
condition (\ref{p}), the horizontal velocity of the flow can be determined,%

\begin{equation}
\vec{u}=\frac{1}{\mu}\left[  \vec{\nabla}p\left(  \frac{z^{2}}{2}-hz\right)
+z\vec{\nabla}\sigma\right]  .
\end{equation}
The vertical velocity is then obtained by using the incompressibility condition,%

\begin{equation}
w=-\frac{1}{2\mu}\left(  \nabla^{2}p\left(  \frac{z^{3}}{3}-z^{2}h\right)
-z^{2}\vec{\nabla}p\vec{\nabla}h+z^{2}\nabla^{2}\sigma\right)
\end{equation}

Substituting these expressions into equation (\ref{interface}) and integrating
over $z$, the final form of the interface equation is derived,%

\begin{equation}
\partial_{t}h=\frac{1}{\mu}\vec{\nabla}\left(  \frac{h^{3}}{3}\vec{\nabla
}p-\frac{h^{2}}{2}\vec{\nabla}\sigma\right)  .
\end{equation}

\section{Numerical Integration Method}

While simple explicit Euler methods are frequently employed in numerical
investigations of nonlinear reaction-diffusion models, such methods easily
become numerically unstable when hydrodynamical microfluidics equations are
considered. Therefore, a specially constructed numerical integration method
has been employed in our study. Below, its brief description is provided.

There are two nonlinear differential equations to solve, one for the thickness
$H$ and the other for the concentration $C$ of the surfactant. We define
vector $\vec{U}=(H,C)$ and formally write both equations as%

\begin{equation}
\partial_{t}\vec{U}=\mathbf{F}(\vec{U})
\end{equation}
The matrix operator $\mathbf{F}$ can be decomposed as $\mathbf{F=L}\vec
{U}+\mathbf{N}(\vec{U})$ into its linear ($\mathbf{L}\vec{U}$) and nonlinear
($\mathbf{N}(\vec{U})$) parts. They are defined as%
\begin{equation}
\mathbf{L=}\left.  \frac{\delta\mathbf{F}(\vec{U})}{\delta\vec{U}}\right\vert
_{\vec{U}=\vec{U}_{0}},\text{ \ }\mathbf{N}(\vec{U})=\text{\ }\mathbf{F}%
(\vec{U})-\mathbf{L}\vec{U}.
\end{equation}

To compute $\vec{U}(t+\Delta t)$ at the next time step $t+\Delta t$, the mixed
semi-implicit method is employed:%

\begin{equation}
\vec{U}(t+\Delta t))-\Delta t\mathbf{L}\vec{U}(t+\Delta t)=\vec{U}(t)+\Delta
t\mathbf{N}(\vec{U}(t)).
\end{equation}
Thus, the implicit method is used to determine the contribution from the
linear part in time $t+\Delta t$ and the explicit method is employed to
compute the nonlinear part contribution.

Applying the inverse matrix operator $(1-\Delta t\mathbf{L})^{-1}$ to both
sides of this equation yields%

\[
\vec{U}(t+\Delta t)=(1-\Delta t\mathbf{L})^{-1}\left(  \vec{U}(t)+\Delta
t\mathbf{N}(\vec{U}(t))\right)
\]

Recombining some terms and using $\mathbf{F=L}\vec{U}+\mathbf{N}(\vec{U})$,
the final finite-difference equation is obtained

\[
\vec{U}(t+\Delta t)=\vec{U}(t)+\Delta t(1-\Delta t\mathbf{L})^{-1}%
\mathbf{F}(\vec{U}(t))
\]

The calculation of the inverse differential matrix operator $(1-\Delta
t\mathbf{L})^{-1}$ is most conveniently performed by transforming the equation
to the Fourier space. Here, it is important that the term $\mathbf{F}(\vec
{U}(t))$ is determined before applying the Fourier transformation.

Our simulations using this numerical method have been performed only in the
one-dimensional case, $H=H(x,t)$ and $C=C(x,t)$. In two dimensions, the method
becomes complicated and such simulations have not been undertaken.

\end{document}